# Modified Micropipline Architecture for Synthesizable Asynchronous FIR Filter Design

Basel Halak and Hsien-Chih Chiu,
ECS, Southampton University, Southampton, SO17 1BJ, United Kingdom
Email: {bh9, hc13g09} @ecs.soton.ac.uk}

**Abstract**   The use of asynchronous design approaches to construct digital signal processing (DSP) systems is a rapidly growing research area driven by a wide range of emerging energy constrained applications such as wireless sensor network, portable medical devices and brain implants.  The asynchronous design techniques allow the construction of systems which are samples driven, which means they only dissipate dynamic energy when there processing data and idle otherwise. This inherent advantage of asynchronous design over conventional synchronous circuits allows them to be energy efficient. However the implementation flow of asynchronous systems is still difficult due to its lack of compatibility with industry-standard synchronous design tools and modelling languages.  This paper devises a novel asynchronous design for a finite impulse response (FIR) filter, an essential building block of DSP systems, which is synthesizable and suitable for implementation using conventional synchronous systems design flow and tools. The proposed design is based on a modified version of the micropipline architecture and it is constructed using four phase bundled data protocol. A hardware prototype of the proposed filter has been developed on an FPGA, and systematically verified. The results prove correct functionality of the novel design and a superior performance compared to a synchronous FIR implementation. The findings of this work will allow a wider adoption of asynchronous circuits by DSP designers to harness their energy and performance benefits.

## 1. Introduction

Asynchronous circuits employ handshake protocols to communicate with their environment, sequence operations and co-ordinate signals transfer. This is a fundamentally different approach from synchronous systems which require a common discrete notion of time (i.e. clock signal) to perform the above tasks. This inherent property gives asynchronous systems many interesting properties, namely: low electromagnetic noise emission,  robustness towards variations in supply voltage, temperature, and fabrication process parameters, improved performance [1]. As such they are widely employed in modern ICs for designing clock generators, building elastic pipelines, or for constructing  secure systems [2].

The use of asynchronous circuits in digital signal processing (DSP) systems is a growing area of research driven by a range of emerging energy constrained applications such as wireless sensor network, portable medical devices and brain implants.

Asynchronous design methodologies are especially relevant for applications where in the system is sampling signals from the environment such as temperature, pressure, electro-cardiograms or speech. Some of these signals evolve smoothly (e.g. temperature); others change sporadically (e.g. speech). However, all of them usually stay constant for relatively long periods of time. This means there is no need to sample such signals on regular basis. This characteristic of such signals has stimulated the development of irregular sampling asynchronous DSP systems. The latter have been demonstrated to achieve significant saving in power consumption [3, 4, 5 and 6].

In addition, the benefits of asynchrony in achieving high performance low latency DSP systems have been highlighted by a number of recent papers [7, 8].  However, asynchronous design methodologies have not yet been widely adopted due to their lack of compatibility with industry standard IC design tools and modelling languages. Existing tools for asynchronous circuit design require a significant re-education of designers, and their capabilities are limited compared to commonly used synchronous tools [9, 10].  One solution to this problem is to develop asynchronous circuits' architectures which can be implemented using the conventional synchronous design tools and flow. One such attempt is presented in [11], where in the authors introduced a synthesisable RTL design for an asynchronous FIFO. More recently the authors of [12] presented  a latch-based design for the C-element.  To the best of our knowledge no previous attempts have been made to construct asynchronous sample driven DSP functions using main stream synchronous design tools and languages. This work reveals for the first time a novel asynchronous architecture of FIR filter, a key component for constructing DSP systems. The design is  based on micropipelines architecture [1], It is  modelled at the register transfer level (RTL) using the HDL language System Verilog, and implemented on an FPGA device.

To validate the functional correctness of the design, it is compared with an equivalent a synchronous FIR filter. Details of the architecture of the proposed design and design validation are presented in the next two sections.

## 2. The Proposed FIR Filter Architecture

The basic structure of an FIR filter is shown in figure 1, synchronous implementation of this filter typically replaces delay element with clocked registers. In order to develop asynchronous implementation of the above architecture, clocked elements should be replaced by an asynchronous control logic which implements a handshake protocol. This work employs the 4-phase bundled-data protocol illustrated in figure 2, its operation is as follows: the sender releases new data items, than it sets the request signal to a high value. The receiver latches the new data and acknowledges data reception by setting acknowledge signal to a high value, which in turn resets the request signal. Finally the receiver rests the acknowledge signal.

This protocol is typically realised using the micropipeline structure shown in figure 3 [1]. In order to realise the functionality of an FIR filter, both the original 4 phase protocol and its implementation had to be modified as illustrated in figure 4.

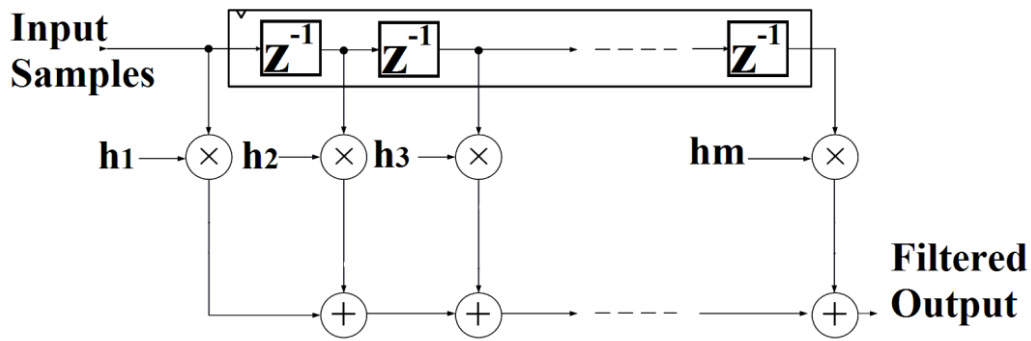

Fig. 1: Structure of FIR filter

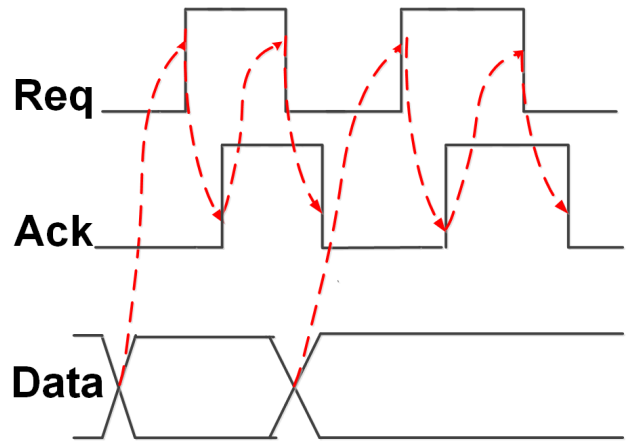

Fig. 2: Four Phase Bundled Data Protocol

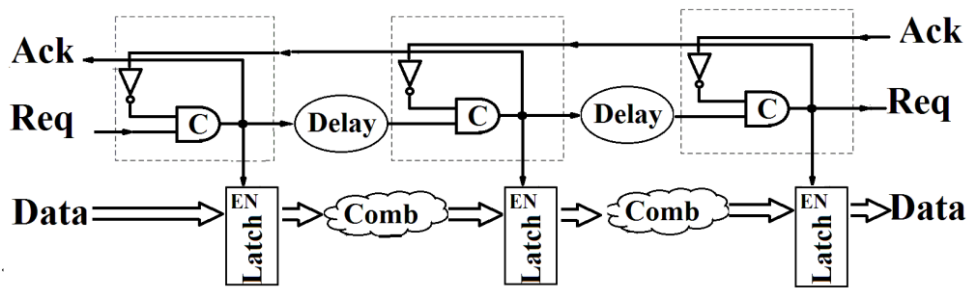

Figure 3: Micropipeline Architecture

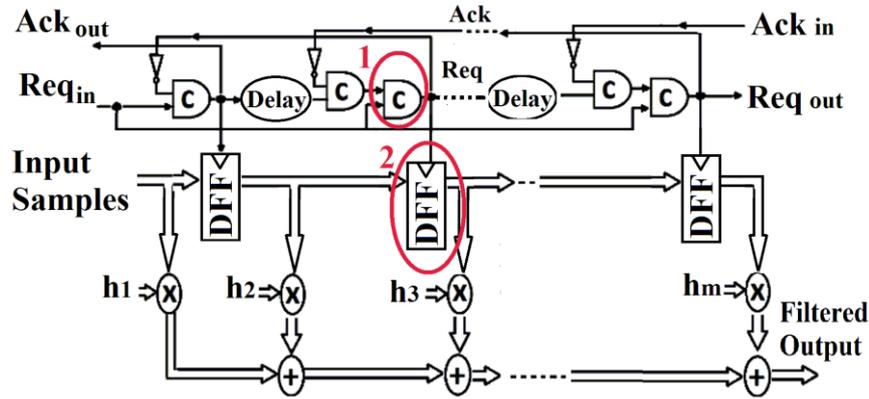

**Fig. 4: The Proposed Asynchronous FIR Architecture**

The micropipeline circuit shown in Figure 2 cannot be used directly in realizing a functional filter this is due to the immediate token occupancy at all stages in the micropipeline upon request, this causes an incoming sample to fill up all elements in the z-domain when they should be holding previous samples. Hence, it is necessary to devise a solution to stop the token from flowing down uncontrollably, while still obeying the communication actions of the 4-phase protocol.

This issue was solved by adding an extra C-element to the original architecture (as highlighted by circle 1 in figure 4). This prevents the token in each stage from propagating down to the next stage. The tokens are temporarily stored at the output of the first C-element until the assertion of the request signal that enters the first stage, i.e. whenever a new sample arrives to the input of the filter along with a request (Req in). This request signal (Req in) can be treated as a "global request" and is gated together with the token of each stage by the second C-element in a stage. The token from each stage is only permitted to fire to the next stage when the global request asserts, where it is again locked down by the second C-element until next assertion of the global request (Req in). This second C-element in each stage also transfers the acknowledgement signal in each stage into global acknowledgement. This is because the 'ack' in each stage asserts now with the global request signal than with the local tokens. The 4-phase handshake actions after modification become:

**Step 1**: The sender sets 'Req' high after the data is ready to be sent and the global request (Req in) is high.
**Step 2**: The receiver receives the data and sets 'Ack' high while the local request (Req in) is still high
**Step 3**: The sender responds to 'Ack' and resets 'Req' that was asserted in Step 1 only when the global request (Req in) becomes low.
**Step 4**: The receiver acknowledges this by resetting 'Ack' that was asserted in Step 2 while the global request (Req in) is still low.

It can be observed that the 4-phase communication actions have now become controlled by the global request signal, which stops the protocol from executing if not set to a high value. This method effectively locks down the token in a stage before new sample arrives at the input of the filter, with just an extra C-element in intermediate stages and in the last stage. It should be noted that the original 4-phase actions shown in figure 2 are still well-preserved.

The inclusion of an additional C-element controlled by the global request signal ensures that the tokens only flow to the next stage upon each arrival of the new sample from the sampling unit in a pipelined manner. However, it can cause a circuit malfunction if level sensitive latches are used in the control logic as in the micropipeline in figure 3. This is due to simultaneous transparency of all data latches as explained below. By considering Step 3 in the modified 4-phase protocol described above, it can be seen that 'Req' will stays asserted as long as the global request signal is still high. Such scenario will cause level sensitive latches in all stages to remain transparent. Consequently, the data item at the input stage of the filter will be copied to all stages of the pipeline and the original data will be corrupted. To tackle this issue, the latches have been replaced by edge-triggered D-type flip-flops (as highlighted by circle 2 in figure 4). This ensures that data in each stage are updated only at the instantaneous moment when Step 1 executes.

## 3. Design Evaluation

To evaluate the proposed architecture a low pass FIR filter has been considered, the design specifications are listed in table 1. A MATLAB tool called "Fdatool" is employed to derive filter's equation. The latter is used to develop two hardware implementations of the filter, one using conventional synchronous structure and the second is based on the proposed asynchronous filter architecture in figure 4. Both filters have been modelled in System Verilog and realised on an Altera Cyclone II FPGA board. The filters have been tested by applying a noisy ECG signal; which was obtained from an online database "Physionet". The chosen data was a raw ECG signal recorded from a male patient at the age of 25 on the date of 7[th] December, 2004. The signal had already been digitized by an ADC with 12-bits resolution at a sampling rate of 125Hz and there were in total 10000 discrete samples. Test results are summarized in Figures 5, 6 and 7. They clearly indicate that the asynchronous filter meets the specification and it is functionally equivalent to its synchronous counterpart.

**Table 1: Low Pass Filter Specification**

| Specification | Sampling Freq. | Pass-band Freq. | Stop-band Freq. | Pass-band Ripple | Stop-band Ripple | Filter Order |
|---|---|---|---|---|---|---|
|  | 125 Hz | 35Hz | 45Hz | 1 dB | 80dB | 32 |

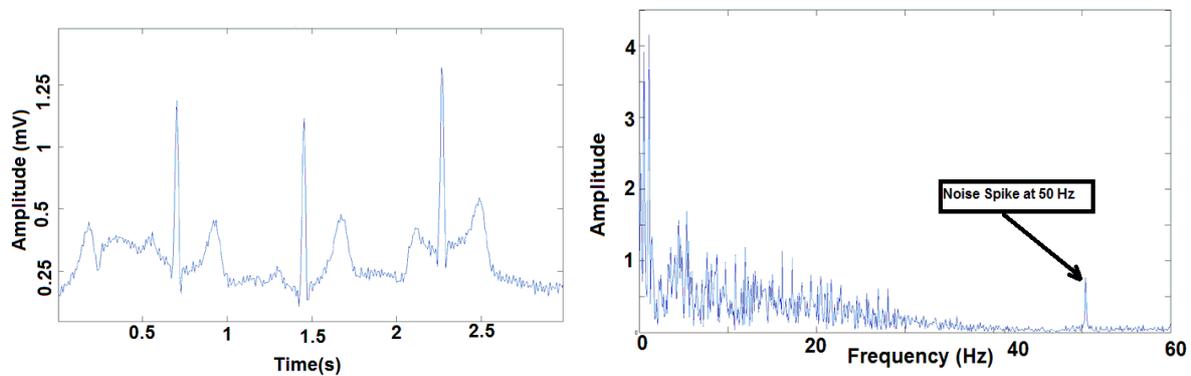

**Fig. 5: Time and Frequency Plots for Unfiltered ECG Signal**

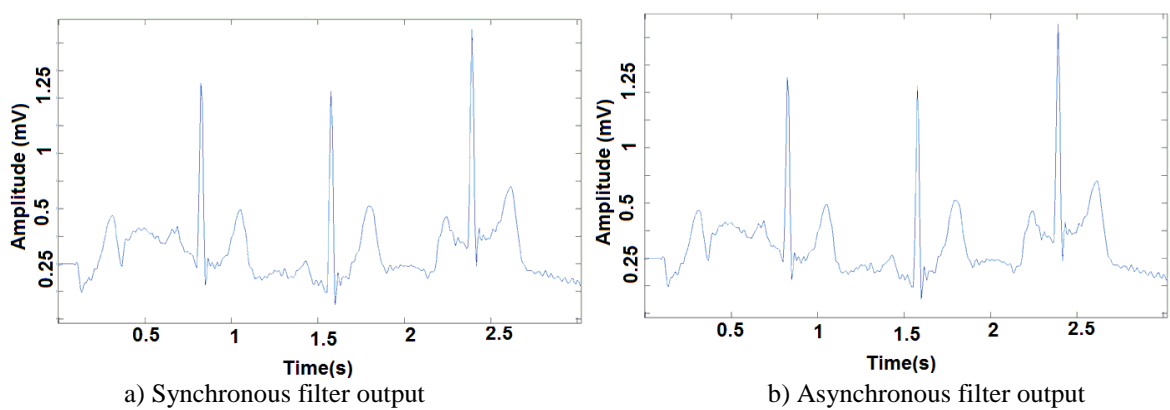

a) Synchronous filter output　　　　　　　　　　b) Asynchronous filter output

**Fig. 6: Output Response Comparison**

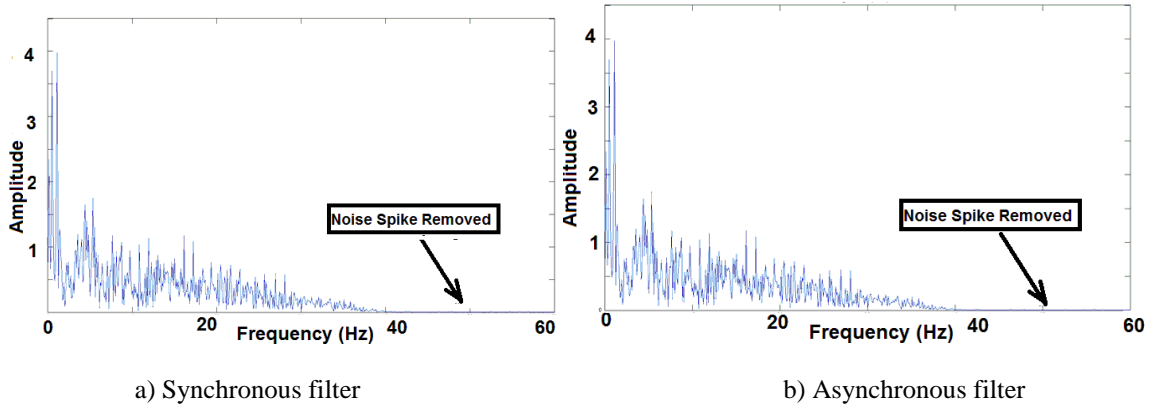

a) Synchronous filter  b) Asynchronous filter

**Fig.7: Frequency Response Comparison**

To compare the proposed asynchronous architecture with the synchronous design, the area and the latency of both implementations have been estimated. The results listed in table 2 indicate that the asynchronous filter has a reduced latency, this is due to the fact that the speed of a synchronous filter is determined by its slowest stage (i.e. critical path delay), so the total latency is a multiple of the delay of the slowest stage, whereas in the asynchronous filter, each stage runs at its own speed, so the total latency is the sum of the delays of all stages. Asynchronous implementation incurs additional area costs (5% in this case), due to the control logic circuitry. A fair comparison of power consumption is more difficult, this is mainly due to the fact that dynamic power is estimated for a specific clock frequency, which is not applicable to a clockless asynchronous design.

**Table 2: FPGA Implementation Area and Performance Comparison**

| Area (Logic Element) | | Latency (ns) | |
|---|---|---|---|
| Asynchronous | Synchronous | Asynchronous | Synchronous |
| 1247 | 1184 | 7.4 | 9.6 |

To illustrate the benefits of the proposed design, a comparison with related work in the literature is presented in Table 3. The use of asynchronous digital filters has been suggested by a number of papers and for different applications, therefore the target application and the implementation technology is included in this comparison for completeness. The figures in Table 3 are extracted directly from the cited papers. It can be noticed that the proposed asynchronous filter exhibits superior performance compared to previous design in [5] and [6] which target similar applications. What is more important is the fact that the proposed design is suitable for both ASIC and FPGA platforms, this is because it is can be modelled using standard HDL languages and implemented using industry standard design tools already used for synchronous designs. This is in contrast with previously proposed asynchronous filter architectures [5], [6] and [7], the latter are only suitable for ASIC platform and can only be designed using custom design tools, which significantly limit their usability and reproducibility.

**Table 3: Asynchronous Digital Filters Comparison**

| Design | Target Application | Sample Frequency (Hz) | Technology (nm) | Implementation | Chip Area (mm2) | Power Supply (V) | Throughput (Bit/S) |
|---|---|---|---|---|---|---|---|
| Xiaofei et. Al. [5] | EEG/ECG Signal Filtering | 2.5 K | 130 | ASIC | N/A | 0.17 | 40 K |
| Singh et. al. [7] | Optical Disc Drive | 0.9G | 180 | ASIC | 0.455 | 1.8 | 14.4 G |
| Nielsen et al. [6] | Hearing Aids | 20 K | 700 | ASIC | 19.36 | 1.55 | 320 K |
| This Work | ECG Signal Filtering | 125 K | 32 | ASIC/ FPGA | 0.01 | 1 | 2 M |

## 4. Conclusions

A new synthesisable architecture for an asynchronous sample driven FIR filter has been developed. The latter is suitable for implementation using industry-compatible IC design tools and flow. The proposed design has been implemented on an FPGA device and verified through rigorous tests. The same design approach can be easily applied to any digital signal processing block, which greatly enhances the ability of IC designers to construct energy efficient sample driven DSP system using asynchronous circuits.

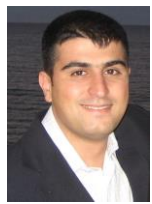
**Basel Halak** received the B.S. degree from the School of Electronics Engineering, Damascus University, Damascus, Syria, in 2001, and the M.Sc. and Ph.D. degrees in microelectronics system design from Newcastle University, Newcastle upon Tyne, U.K., in 2005 and 2009, respectively. He joined the University of Southampton, Southampton, U.K., in 2011. He has authored a monograph and more than 30 refereed papers. His current research interests include dependable systems on a chip, fault tolerance techniques, and VLSI circuits for communications.

**Hsien-Chih Chiu** received his master degree in system-on-chips from the University of Southampton in 2013 then join ARM ltd where he is currently working as a design engineer.